\documentclass[conference]{IEEEtran}
\IEEEoverridecommandlockouts
\usepackage{cite}
\usepackage{amsmath,amssymb,amsfonts}
\usepackage{algorithm}
\usepackage{algpseudocode}
\usepackage{graphicx}
\usepackage{textcomp}
\usepackage{xcolor}
\usepackage{enumitem}
\def\BibTeX{{\rm B\kern-.05em{\sc i\kern-.025em b}\kern-.08em
    T\kern-.1667em\lower.7ex\hbox{E}\kern-.125emX}}
\begin{document}

\title{Intelligent Duty Cycling Management and Wake-up for Energy Harvesting IoT Networks with Correlated Activity} 
%A Holistic Approach to Sustainable IoT Development: Integrating Energy-Efficient Technologies and Intelligent Solutions
%{\footnotesize \textsuperscript{*}Note: Sub-titles are not captured in Xplore and should not be used}

\author{\IEEEauthorblockN{1\textsuperscript{st} David E. Ru\'{i}z-Guirola}
\IEEEauthorblockA{\textit{Centre for Wireless Communications} \\
\textit{University of Oulu}\\
Oulu, Finland \\
David.RuizGuirola@oulu.fi}
\and
\IEEEauthorblockN{2\textsuperscript{nd} Onel L. A. L\'{o}pez}
\IEEEauthorblockA{\textit{Centre for Wireless Communications} \\
\textit{University of Oulu}\\
Oulu, Finland \\
Onel.AlcarazLopez@oulu.fi}
\and
\IEEEauthorblockN{3\textsuperscript{rd} Samuel Montejo-S\'{a}nchez}
\IEEEauthorblockA{\textit{Instituto Universitario de Investigaci\'{o}n y} \\
\textit{Desarrollo Tecnol\'{o}gico} \\
\textit{Universidad Tecnol\'{o}gica Metropolitana}\\
Santiago, Chile \\
smontejo@utem.cl}
\and
\IEEEauthorblockN{4\textsuperscript{th} Israel Leyva Mayorga
}
\IEEEauthorblockA{\textit{Department of Electronic Systems} \\
\textit{Aalborg University}\\
Aalborg, Denmark\\
ilm@es.aau.dk}
\and
\IEEEauthorblockN{5\textsuperscript{th} Zhu Han
}
\IEEEauthorblockA{\textit{Department of Electrical} \\
\textit{and Computer Engineering} \\
\textit{University of Houston}\\
Houston TX, USA\\
zhan2@central.uh.edu
}
\and
\IEEEauthorblockN{6\textsuperscript{th} Petar Popovski
}
\IEEEauthorblockA{\textit{Department of Electronic Systems} \\
\textit{Aalborg University}\\
Aalborg, Denmark\\
petarp@es.aau.dk
}

\thanks{%David E. Ru\'{i}z-Guirola and Onel L. A. L\'{o}pez are with the Centre for Wireless Communications, University of Oulu, Finland. \{{David.RuizGuirola, Onel.AlcarazLopez\}.  
%Samuel Montejo-S\'{a}nchez is with the {Instituto Universitario de Investigaci\'{o}n y Desarrollo Tecnol\'{o}gico, Universidad Tecnol\'{o}gica Metropolitana}, Santiago, Chile. \{smontejo@utem.cl\}.
This work has been partially supported %Chile by ANID FONDECYT Iniciaci\'on No. 11200659, %SCC-PIDi-UTEM 
%FONDEQUIP-EQM180180, and Collaborative Research Activities between PIDi/UTEM and FIE/UCLV, in Brazil by CNPq (402378/2021-0, 305021/2021-4), Print CAPES-UFSC ``Automation 4.0'', and RNP/MCTIC (Grant 01245.010604/2020-14) 6G Mobile Communications Systems, and in 
by the Finnish Foundation for Technology Promotion and the Research Council of Finland (former Academy of Finland) 6G Flagship Programme (Grant Number: 346208), the Finnish Foundation for Technology Promotion, and by the European Commission through the Horizon Europe/JU SNS project Hexa-X-II (Grant Agreement no. 101095759), and in Chile by ANID FONDECYT Regular 1241977.  
%Manuscript received April 19, 2005; revised August 26, 2015.
}%}
}

\maketitle

\begin{abstract}
%This paper presents a comprehensive approach to sustainable development in the domain of IoT (Internet of Things). It integrates energy-efficient technologies, optimized resource allocation strategies, and intelligent solutions driven by machine learning (ML). The aim is to manage the resources in an IoT environment efficiently, so as to prolong the battery life of devices and reduce instances of low-energy availability. 
%Herein, a sustainable IoT scenario with energy harvesting capabilities is modelled using a four-state Markov chain to represent the states of IoT devices (IoTDs), and modulated Poisson processes for energy harvesting availability. The battery state is modelled as a semi-Markov chain. An optimization problem is formulated considering the energy efficiency trade-offs, spatial and temporal correlations between IoTDs. Moreover, we propose a duty cycling configuration solution based on K-nearest neighbors (KNN). 
%The results demonstrate significant improvements in energy savings and performance, with up to 11 times less misdetection probability and a 50\% reduction in energy consumption observed in high-density scenarios.
This paper presents an approach for energy-neutral Internet of Things (IoT) scenarios where the IoT devices (IoTDs) rely entirely on their energy harvesting capabilities to sustain operation. We use a Markov chain to represent the operation and transmission states of the IoTDs, a modulated Poisson process to model their energy harvesting process, and a discrete-time Markov chain to model their battery state. The aim is to efficiently manage the duty cycling of the IoTDs, so as to prolong their battery life and reduce instances of low-energy availability. We propose a duty-cycling management based on K- nearest neighbors, aiming to strike a trade-off between energy  efficiency and detection accuracy. This is done by incorporating spatial and temporal correlations among IoTDs' activity, as well as their energy harvesting capabilities. We also allow the base station to wake up specific IoTDs if more information about an event is needed upon initial detection. Our proposed scheme shows significant improvements in energy savings and performance, with up to 11 times lower misdetection probability and 50\% lower energy consumption for high-density scenarios compared to a random duty cycling benchmark. 
\end{abstract}

\begin{IEEEkeywords}
Duty cycling, energy harvesting, Internet of Things, K-nearest neighbors, wake-up. 
\end{IEEEkeywords}

\section{Introduction}\label{intro}

%Pursuing sustainable development in today's society requires technological solutions that can effectively balance economic growth, social equity, and environmental integrity. The Internet of Things (IoT) is critical to achieving sustainability as it can improve urban management by connecting devices for mobility, safety, environmental, and resource monitoring~\cite{lopez2023energy,belli2020iot}. 
%The sixth generation (6G) of wireless systems need to support massive connectivity while delivering high performance~\cite{mahmood2020white}. In networks with a high density of devices, achieving sustainability becomes even more pressing due to the growing complexity of the network, increasingly stringent service requirements, and the often-challenging features of devices, particularly energy availability in low-power IoT deployments. 
%The exponential growth of IoT devices (IoTDs) aggravates these challenges, highlighting the urgent need for solutions to minimize energy consumption and enhance resource utilization~\cite{lopez2023energy,qin2019low}. 
%Additionally, there are environmental concerns surrounding device manufacturing and disposal when replacing batteries~\cite{benkhelifa2020recycling}.

Achieving sustainable development in today's society requires technological solutions balancing economic growth, social equity, and environmental integrity. In this context, the Internet of Things (IoT) plays a crucial role %in achieving sustainability 
by connecting devices for mobility, safety, environmental, and resource monitoring~\cite{lopez2023energy,belli2020iot}. 
%The of wireless systems must support massive connectivity while delivering high performance~\cite{mahmood2020white}. 
In the sixth generation (6G) networks, achieving sustainability becomes even more critical due to the huge IoT device (IoTD) expected density, highlighting the urgent need to minimize IoTDs' energy consumption and enhance resource utilization to deliver high performance~\cite{qin2019low}. 
To tackle sustainability issues in IoT ecosystems, it is crucial to rely less on external power sources and instead use energy harvesting (EH) techniques. This, %will enhance their energy autonomy. Additionally, it is important to optimize 
together with 
resource allocation optimization for energy efficiency~\cite{lopez2021survey}, promotes energy autonomy. %Therefore, intelligent resource management plays a key role in achieving maximum energy efficiency~\cite{lopez2021survey}. 
Several techniques such as wake-up receivers (WuRs) and discontinuous reception (DRX) %duty cycling 
are being explored to increase the battery life of devices and reduce energy waste. 
Specifically, WuRs have the potential to save up to 1,000 times~\cite{ruiz2022energy} the energy consumed by the main radio interface, allowing IoTDs to transition to a sleep state while the WuR remains active~\cite{petar1, petar2}. 
However, these methods may not always yield the desired results, owing to the irregular traffic patterns of IoTDs~\cite{ruiz2022energy}. 
Also, IoTDs relying solely on ambient EH sources for charging suffer from uncertain energy availability, jeopardizing their ability to recharge a limited battery and remain operational. This makes it difficult to coordinate the amount of energy available in each device and manage the duty cycling accordingly to extend the network's lifetime, scalability, and availability for proper functioning, and to ensure that the network is self-sustainable during its lifetime. Monitoring the network and keeping track of the IoTDs' traffic generation, and incoming energy availability simultaneously is an extremely challenging task~\cite{lopez2023energy}.    
In this sense, 
machine learning (ML) techniques have proven to characterize well the spatio-temporal nature of IoT traffic generation. ML can optimize resource allocation, predict system changes, and enhance energy efficiency in IoT networks~\cite{ul2022learning}. 
%Additionally, there has been recent interest in stochastic EH models where the processes of energy renewal are considered to be random~\cite{ku2015advances}. One significant advantage of this type of model is the simplicity and the fact that there is no requirement for non-causal knowledge of energy state information. 
However, despite the benefits, ML-based methods for adjusting WuR and duty cycling parameters in self-sustaining IoT networks have not been explored, offering potential for further energy efficiency.

\begin{figure}[t]
\centerline{\includegraphics[width=0.7\columnwidth]{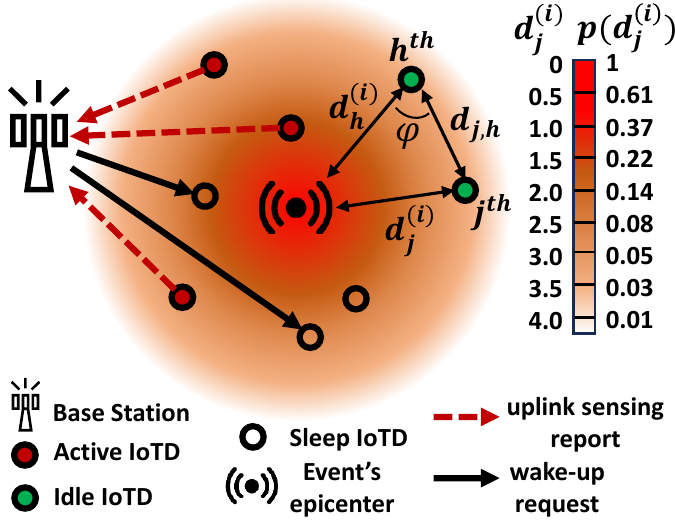}}
\caption{Illustration of an IoT network where the BS controls and collects information from various IoTDs. The impact of an event on the surrounding IoTDs is modeled through a probability activation function that decays with the distance (meters) from the event epicenter to the IoTDs.}
\label{fig1}
\end{figure}

%Overall, sustainability in IoT can be developed by using ultra-low power WuR architecture, duty cycling management, and EH. A comprehensive approach integrating these technologies with intelligent solutions can advance IoT ecosystems towards a sustainable future. By using ML-based methods to adjust WuR and duty cycling parameters, energy consumption can be further reduced.
%In a self-sustainable scenario, as the one presented in Fig.~\ref{fig1}, in which EH-powered devices are waiting to be triggered by measurements of an event, is paramount to implement energy-efficient mechanisms that lower the IoTD's energy consumption while allowing the IoTDs to alternate in a duty cycling between idle/sensing and sleep states to guarantee the IoTDs EH process while in sleep state aiming for high energy availability for the IoTDs to operate. 
%Herein, efficient and sustainable IoT deployment can be achieved through a combination of low-power WuR architecture, duty cycling configuration, and EH. 
%Thus, a comprehensive approach integrating these technologies with intelligent solutions is necessary for a sustainable IoT future. 
In several IoT scenarios, IoTDs are deployed to sense/detect events/alarms, as depicted in Fig.~\ref{fig1}. 
Aiming for self-sustainable IoT deployments, these IoTDs may rely on energy harvested from the environment and an efficient energy consumption to remain operational. To ensure this, it is crucial to implement energy efficient mechanisms that allow the IoTDs to alternate between idle/sensing and sleep states. 
This duty cycling mode guarantees high energy availability for operation. Achieving a sustainable IoT deployment requires the integration of low-power WuR architecture, energy efficient duty cycling configuration, and efficient EH. 
Therefore, a comprehensive approach that integrates these technologies with intelligent solutions is essential for a sustainable IoT deployment. 
In this paper, we aim to efficiently manage the duty cycling in an energy-neutral IoT scenario, extend the battery life of IoTDs, and use WuR to avoid misdetected events while avoiding low-energy availability. 
Specifically, the contributions of this paper are
summarized as follows:  
%To achieve this, 
\begin{itemize}
    %\item 
    \item We address an energy-neutral IoT scenario where the IoTDs harvest ambient energy and monitor the occurrence of events/alarms while in the sensing state. Upon the detection of an event/alarm, they transmit this information to the BS. We employ a four-state Markov chain to represent the  operation and transmission states of the IoTDs. The energy harvesting availability is modeled as modulated Poisson processes (MPP), and the battery state as a discrete Markov chain.   
    %\item We model the energy harvesting availability as modulated Poisson processes (MPP), and the battery state as a semi-Markov chain. 
    %Furthermore, 
    \item We propose a K-nearest neighbors (KNN)-based duty cycling configuration that considers both energy efficiency and the likelihood of event misdetection. This perspective considers spatial and temporal correlations among IoTDs. 
    %Additionally, 
    \item We propose a wake-up method for WuR-equipped IoTDs. The BS can wake up certain IoTDs if more information about the event is needed. Herein, the wake-up process is managed based on the spatial correlation among IoTDs. 
    %Finally, 
    \item We show the improvement in energy saving and performance from our proposal with up to 11 times less misdetection probability and 50\% decrease in energy consumption for high-density scenarios compared to a random duty cycling baseline. 
\end{itemize} 
To the best of our knowledge, optimizing duty cycling in this type of scenario has not been proposed before. Our research is particularly important as the number of IoTDs rises, making essential to reduce their energy consumption to meet operational requirements and achieve sustainable IoT networks.

The rest of the paper is organized as follows. In Section~\ref{system}, we present the system model along with the analysis of EH and event influence. Section~\ref{analisis} outlines the optimization problem for the duty cycling setup, while Section~\ref{proposal} proposes solutions based on heuristics and ML. The effectiveness of the proposed methods is evaluated through numerical simulations in Section~\ref{result}. Finally, we conclude the paper in Section~\ref{conclusion}.

\section{System Model}\label{system}

\begin{figure}[t]
\centerline{\includegraphics[width=0.8\columnwidth]{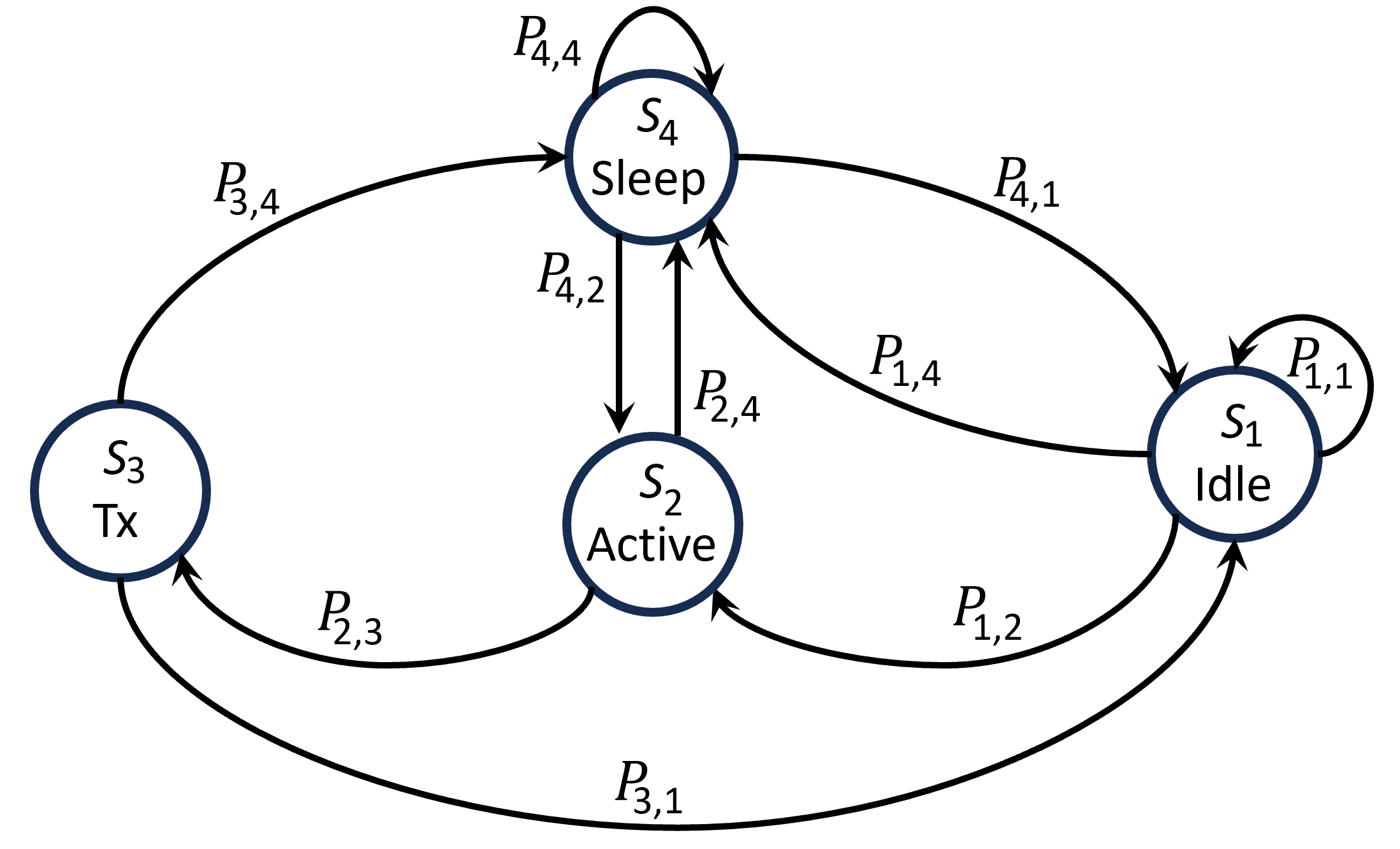}}
\caption{Device's operation states modeled as a four states discrete-time Markov chain.}
\label{fig2}
\end{figure}

We consider the coverage area of a BS serving as gateway for a set ${\mathcal{N}}$ of $N$ IoTDs with EH capabilities as depicted in Fig.~\ref{fig1}. These IoTDs are deployed to identify event triggers, such as a moving object in motion detection applications or a fire in a fire-alarm system. 
Moreover, the time is slotted in transmission time intervals (TTIs) with duration $\tau$. 
%time is divided into Transmission Time Intervals (TTIs) with duration $\tau$,
%we assume time is slotted in transmission time intervals (TTI). 
We assume that the position of each IoTD is known by the coordinator and that events occur uniformly in the area with probability $\alpha$ in each TTI~\cite{ruiz2022energy}. %Once the event is detected, the activation of the detecting IoTDs is modeled as a four-state Markov chain as illustrated in Fig.~\ref{fig2}. 
The IoTDs can operate in four different states $S_m \in {\mathcal{S}}$, with $m \in [1,2,3,4]$:
\begin{enumerate}%[label=(\roman*)]
    \item idle state ($S_1$), wherein the IoTDs wait to be triggered by an event; 
    \item active state ($S_2$), wherein the IoTD is triggered by an event and aims to send the information to the BS;
    \item transmission state ($S_3$), wherein the IoTD sends the information to the BS;
    \item sleep state ($S_4$), wherein the IoTD stays at low-power state and harvests energy.
\end{enumerate}
%(i) idle state ($S_1$) where the IoTDs wait to be triggered by an event; (ii) active state ($S_2$), the IoTD is triggered by an event and aims to send the information to the BS; (iii) transmission state ($S_3$), the IoTD sends the information to the BS; and (iv) sleep state ($S_4$) when the IoTD stays at low-power state and harvests energy.  
We denote $P_{m,n}^{(j)}$ as the transition probability for IoTD $j$ from a state $S_m$ into a state $S_n$.  %where $S_m, S_n \in  {\mathcal{S}}$.   
The states and their transition probabilities are illustrated with a discrete-time Markov process, as shown in Fig.~\ref{fig2}. 

A device goes from $S_1$ to $S_2$ with probability $P_{1,2}^{(j)}$ upon detecting an alarm event, indicating the need for data transmission to the BS. IoTDs in $S_2$ with high-energy availability transmit the information corresponding to the detected events according to their battery level with probability $P_{2,3}^{(j)}$, while those with low-energy availability cannot transmit the information and go to $S_4$ awaiting to harvest enough energy for future events. Notice that an IoTD with high-energy availability can alternate between $S_1$ and $S_4$ with probabilities $P_{4,1}^{(j)}$ and $P_{1,4}^{(j)}$ based on their duty cycling configuration and energy availability. 

%We assume that this transitions from sleep to idle state are controlled by the BS according to its knowledge of the network and about the energy availability of the IoTDs. While, the IoTDs are equiped with a wake-up radio (WuR) module to detect the BS request to pass to active state.  Then, using this WuR module aim to save the remaining energy in IoTDs with low-energy availability while those harvest energy and at the same time are available in case the BS decides to request any additional information from an event.    

We define $p(d_{j}^{(i)})$ as a sensing power function for IoTD $j\in \mathcal{N}$ while in $S_1$, representing the impact of the $i^{th}$ event with  epicenter at $(x_i, y_i)$, within the network area $\xi \subset \Re^{2}$. Here, $d_{j}^{(i)}$ indicates the distance between them. Note that, $p(d_{j}^{(i)}): [0, \infty) \rightarrow (0,1]$ is non-increasing to model a decaying influence of events as the distance $d_{j}^{(i)}$ increases. Fig.\ref{fig1} illustrates an example of the influence of an event epicenter on the surrounding IoTDs. %Various functions satisfy the stated conditions, such as exponentially decreasing\cite{thomsen2017traffic,FWuS}, linear~\cite{alves2021wireless}, piece-wise linear~\cite{hejselbaek2018empirical,yang2020resource}, or power-law decay~\cite{alves2021wireless} for general scenarios. For more stringent scenarios like indoor IIoTD, step functions~\cite{sun2019modeling} or sigmoidal models~\cite{alves2021wireless} may be employed. 
In this paper, we assume an exponentially decreasing function, i.e., $p(d_{j}^{(i)}) = e^{-\eta d_{j}^{(i)}}$~\cite{ruiz2022energy}. Here, the parameter $\eta$ ($\eta > 0$) controls the average sensitivity for a given distance. 
Furthermore, we assume that the temporal correlation with the event is instantaneous while the spatial correlation depends on $p(d_{j}^{(i)})$.  
Active devices transmit packets to the BS, managing all information exchanges within its cell~\cite{ruiz2022energy}. Assume equal transmission power for all IoTDs. %time is divided into Transmission Time Intervals (TTIs) with duration $\tau$, where 
In a TTI $k$ and state $S_2$ (active), devices generate traffic, while no traffic is generated in the other states.

Active IoTDs send information about the detected event. The level of information sent by each device corresponds to the detected entropy of the event itself. In this study, we model this relationship of detected information using the sensing power function of each IoTD in $S_2$ as follows: 
\begin{equation}
    I_{j}^{(i)} = \Psi p(d_{j}^{(i)}),
    \label{entropia}
\end{equation}
where $\Psi > 0$ constitutes the maximum amount of informacion that an IoTD can detect. %, and $\sigma_j$ is the active factor given by 
%\begin{equation}
    %\begin{array}{rl}
    %    \sigma_j &= \begin{cases}
    %    1,   &\text{if IIoTD $j^{\text{th}}$ is active,}\\   
    %    \hfil 0,   &\text{otherwise,}\\
    %\end{cases}\\
    %\end{array}
%\end{equation}
%Therefore, the information collected by the BS from an event is then measured by the collective information from all active IoTDs. 
%The closer to the event epicenter, the greater the information. 
The BS gathers the information reported by the IoTDs while keeping the one that best describes the event $(i)$. Therefore, the information  about event $i$ at the BS is given by 
\begin{equation}
    %I = \sum_{j=1}^{\mathbb{\mathcal{N}}} I_{j}
    %I = \max([I_{j}, I_{2}, \cdots, I_{\mathbb{\mathcal{N}}}]).
    I^{{(i)}} = \max(I_{j}^{(i)}), \forall j \text{ in } S_2.
    \label{totalentropia}
\end{equation}
If no IoTD detects the event, then $I^{{(i)}} = 0$. 

\begin{figure}[t]
\centerline{\includegraphics[width=\columnwidth]{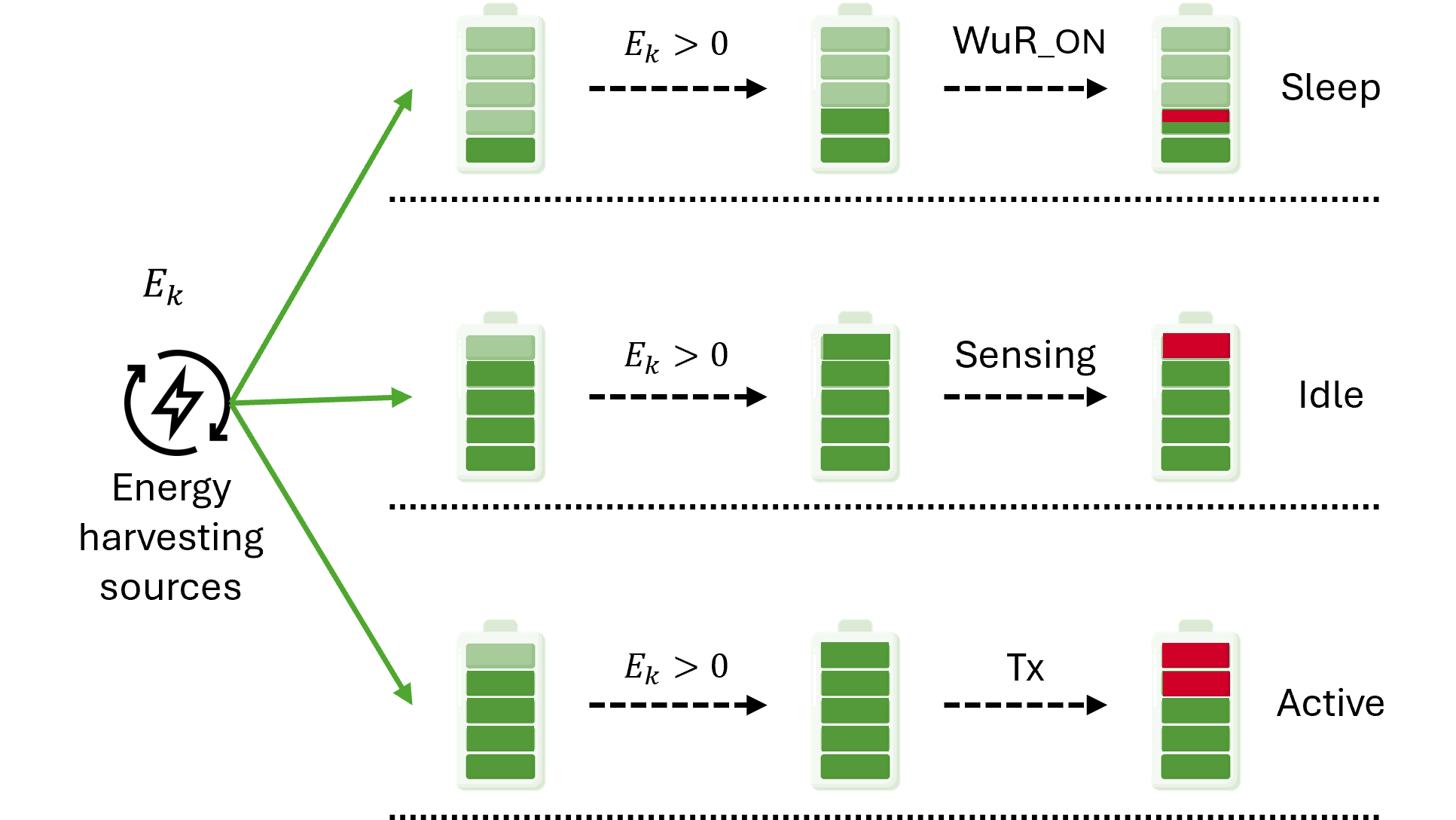}}
\caption{Illustration of the battery level evolution by considering the energy consumption and EH models. Transmission (TX) results in the highest energy depletion, while sleep is a low-energy consumption state. Sensing results in a depletion smaller than TX but greater than in a sleep state.}
\label{fig3}
\vspace{-3mm}
\end{figure}

%Based on the discrete-time approach, the time elapsed to harvest and/or consume energy is discretized with a time slot, transmission time interval (TTI). 
Our model uses discrete energy harvesting and storage for device batteries, which have maximum capacity $E_{\text{max}}$ when fully charged. %Meanwhile transmission in $S_2$ incurs in $E_\text{Tx}$ units of depletion and monitoring the events in $S_1$ in $E_\text{idle}$ units per TTI. 
Additionally, transmitting in $S_2$ results in a depletion of $E_\text{Tx}$ units, while monitoring events in $S_1$ consumes $E_\text{idle}$ units per TTI, as depicted in Fig.~\ref{fig3}. 
We assume that, if the battery level ($B$) of an IoTD at the moment of transmissions is lower than the level needed to transmit or perform event monitoring, the battery is depleted and the action is unsuccessful. We also assume that the BS knows or at least predicts the battery level of the devices according to the information received from those IoTDs in $S_3$. This approach helps to conserve power by not prompting a device with low energy levels to wake up and transmit information. Such an action would otherwise drain the battery, and even if the device were to transmit, the information would not be received due to the lack of energy required for transmission.  

In this paper, we adopt a simplified model based on~\cite{ruiz2022energy} that refers only to the energy consumption of the radio frequency interface. Specifically, the mean energy consumption ($\overline{\text{EC}}$) is calculated based on the energy consumption of battery units ($\text{EC}_{m}$) for every state $S_{m}$. This is summarized as the energy units consumed per IoTD per TTI and written as  
\begin{equation}
    %\overline{\text{EC}} \!=\! \frac{\sum\limits_{j=1}^{\mathcal{N}}\! \sum\limits_{i=1}^{4}\! \Pr({S_i})t_{i}\text{EC}_{i}\!}{\mathcal{N} \sum\limits_{i=1}^{4} \Pr({S_i})t_{i}}.
    \overline{\text{EC}} \!=\! \frac{1}{N}{\sum\limits_{j=1}^{{N}}\! \sum\limits_{m=1}^{4}\! \Pr({S_m^{(j)}})\text{EC}_{m}\!}. 
\label{EC}	
\end{equation}
%where $N$ is the amount of IoTDs. 
%Moreover, 

At each TTI, we also model the energy harvested by the IoTDs by using ${{N}}$ MPPs, one for each IoTD. 
Specifically, a Poisson process is used to describe the occurrence of energy arrival within a certain time interval. The time duration for the occurrence of energy arrival, i.e., the time duration in case the energy source remains active, is modeled with an exponential distribution with  mean $\mu_j$. %Similarly, the exponential distribution with the mean ($\widehat{\mu_e}$) can be used to model the time duration for the absence of energy arrival. 
At each TTI ($k$), the energy arrival from an energy source occurs at maximum only once. In that case, the energy source can be considered as active or inactive within one time slot, as shown in Fig.~\ref{fig3}. By defining the parameters $\lambda_j = 1/\mu_j$, the probability of the energy source being active within one TTI can be expressed as
\begin{equation}
    p_{j} = \lambda_j \tau \exp(-\lambda_j \tau), k = 1, 2,..., K.
    \label{poissonenergy}
\end{equation} 

%\section{Modelling of Analytical State and Energy Harvesting for the Optimization Problem}
\section{Battery State Evolution}

The probability that an IoTD has sufficient energy in the battery to perform transmission or sensing depends on the energy harvested at each TTI plus the initial battery level and previous actions performed by the IoTD. To calculate this probability, we can use a discrete-time Markov chain to model the battery level over time. We define states for the battery level, while transitions occur based on the energy harvesting process and the battery consumption operations. %Then, we'll calculate the probability of being in a state where there is enough battery to perform the operation that needs $E_\text{Tx}$ units. 

Following the MPP, it is possible to harvest energy with a probability $\exp{(-\lambda_j)}$.  
We formulate a discrete Markov chain with states representing the battery level (from 0 to $E_\text{max}$ units). Let's denote the states of the battery level as follows:
\begin{itemize}
    \item State 0: Battery level is 0 units;
    \item State 1: Battery level is $\lambda_j$ units;
    \item State 2: Battery level is $2\lambda_j$ units;
    \item \dots
    \item State $\left\lceil E_\text{max}/\lambda_j \right\rceil$: Battery level is $E_\text{max}$ units. 
\end{itemize}
We calculate the transition probabilities between these states based on the energy harvesting process and battery consumption operations. Assuming a steady state, the state arrival rate is equal to the state departure rate. Therefore, we can use the fact that $\sum_{B = 0}^{\left\lceil E_\text{max}/\lambda_j \right\rceil} r_B = 1$, where $r_B$ is the steady $B$ state probability, to solve the next equation system $P_B \times R_B = C$, where ${C = [1, 1, \cdots, 1]^T}$, 
\begin{align}
    P_B = \begin{pmatrix}
{B_{0,0}} & {B_{0,1}} & \cdots & {B_{0,\left\lceil E_\text{max}/\lambda_j \right\rceil}} \\
{B_{1,0}} & {B_{1,1}} & \cdots & {B_{1,\left\lceil E_\text{max}/\lambda_j \right\rceil}} \\
\vdots & \vdots & \ddots & \vdots \\
{B_{\left\lceil E_\text{max}/\lambda_j \right\rceil,0}} & {B_{\left\lceil E_\text{max}/\lambda_j \right\rceil,1}} & \cdots & {B_{\left\lceil E_\text{max}/\lambda_j \right\rceil,\left\lceil E_\text{max}/\lambda_j \right\rceil}} \\
1 & 1 & \cdots & 1\\
\end{pmatrix}  
    \label{Markov_Battery3}
\end{align}
is the transition matrix, and ${R_B = [r_0, r_1, \cdots, r_{\left\lceil E_\text{max}/\lambda_j \right\rceil}, 1]^T}$. 

Next, we aim to find the probability of being in a state with sufficient battery to perform the operation that needs $E_\text{Tx}$ units, \textit{i.e.,} $\Pr(B\ge E_\text{Tx})$. 
Note that a device goes to state $S_2$ with probability $P_{1,2}^{(j)} + P_{4,2}^{(j)}$. While in $S_2$, we need to check whether the IoTD has sufficient energy to transmit the sensed information.  %can calculate $P_{2,3}^{(j)} = \Pr(B^{(j)}\ge E_\text{Tx})$ by following 
Exploiting 
%the binomial distributions in 
the previously presented Markov chain with states representing $B$, we obtain $Pr(B^{(j)}\ge E_\text{Tx})$ as 
\begin{align}
    &\sum_{x = B_{E_\text{Tx}}}^{B_{\left\lceil E_\text{max}/\lambda_j \right\rceil}} \binom{x}{E_\text{Tx}} (P_{1,2}^{(j)} - P_{4,2}^{(j)})^{E_\text{Tx}} (1 - P_{1,2}^{(j)} - P_{4,2}^{(j)})^{x-E_\text{Tx}},  
    \label{Markov_Battery}
\end{align}
which has a semi-closed form using the regularized incomplete beta function, $\beta_{1 - P_{2,1}^{(j)} - P_{3,1}^{(j)}}(E_{\text{Tx}} + 1, E_\text{max}+1)$,  with parameters $E_{\text{Tx}} + 1$ and $E_\text{max} + 1$~\cite{egorova2023computation}. %This expression represents
%\begin{align}
    %\beta_{1 - P_{1,2}^{(j)} - P_{4,2}^{(j)}}(E_{\text{Tx}} + 1, E_\text{max}+1). 
    %\label{Markov_Battery1}
%\end{align}
We can similarly calculate $\Pr(B^{(j)}\ge E_\text{idle})$, \textit{i.e.}, the probability that the IoTD is able to sense in $S_1$.

\section{On-demand Wake-up and Transition Probabilities}\label{analisis}
%\section{Analytical IoTD State Modeling}

%Here, we first present an analytical model for the device's states aforementioned.  
In this paper, we propose equipping IoTDs with WuR modules to detect requests from the BS to either enter the active state ($S_2$) or switch between states. The WuR module allows IoTDs with low-energy availability to remain in state $S_4$ while also ensuring their availability to provide additional information to the BS when required ($P_{4,2}^{(j)}$).
The BS may try to wake up certain IoTDs in $S_4$ to sense potential events based on information received from other IoTDs and their spatial correlation. However, this might not always be possible because IoTDs with a low-energy availability stay in $S_4$ (with probability $P_{4,4}^{(j)}$) and harvest energy for future events (if enough energy becomes available). Meanwhile, IoTDs in $S_4$ with high-energy availability may be authorized by the BS to move from $S_4$ to $S_1$ for the purpose of monitoring potential events. 

Transitions between states can occur as follows. 

\begin{enumerate}
    \item A transition from $S_4$ to $S_1$ occurs during the ON time of the duty cycling, then $P_{4,1}^{(j)} = \frac{\text{ON time}}{\text{DRX cycle}} $, where DRX cycle accounts for the time between consecutive ON states ($S_1$). Otherwise, the IoTD stays on $S_4$ with probability ${P_{4,4}^{(j)} = 1 - P_{4,1}^{(j)} - P_{4,2}^{(j)}}$. 
    \item A transition from $S_4$ to $S_2$ occurs when the BS needs more information about an event. To determine this transition, we use the conditional probability $\Pr(S_3^{(j)}|S_3^{(h)})$, which represents the probability that an IoTD $j$ is in state $S_3$, given that another IoTD $h$ is in state $S_3$. 
    %$\left(S_3^{(j)}\right)$
    We then assign the maximum conditional probability as ${P_{4,2}^{(j)} =  \max_h\left(\Pr(S_3^{(j)}|S_3^{(h)})\right)}$ with respect to the active IoTDs $h$ in state $S_3$, $\forall h$ in $S_3$. 
    Herein, $\Pr(S_3^{(j)}|S_3^{(h)})$ is determined by the spatial correlation between the $j^{\text{th}}$ IoTD and each $h$ in $S_3$. We calculate $\Pr(S_3^{(j)}|S_3^{(h)})$ using the cosine rule as follows 
\begin{equation}
   {d_{j}^{(i)}}^2 = {d_{h}^{(i)}}^2 + d_{j,h}^2 - 2d_{h}^{(i)}d_{j,h}\cos{\varphi},
\end{equation}
where $d_{j,h}$ the distance between both devices and $\varphi$ is the angle between $d_{h}^{(i)}$ and $d_{j,h}$, as depicted in Fig.~\ref{fig1}. Then, $\Pr\left( S_3^{(j)}|S_3^{(h)} \right)$ is given by  
\begin{equation}
   \exp{\left( -\eta \sqrt{{d_{h}^{(i)}}^2 + d_{j,h}^2 - 2{d_{h}^{(i)}}d_{j,h}\cos{\varphi}} \right)},
\end{equation}
where $d_{h}$, $d_{j,h}$, and $\varphi$ are known. %has a uniform distribution in $2\pi$. %Therefore, $\Pr(A_j | A_h)$ can be written as 
%\begin{equation}
   %\Pr\left( {\varphi} \geq \cos^{-1}\left(\frac{\mathrm{ln^2}(\delta_j)/\eta^2 - d_{h}^2 - d_{j,h}^2}{- 2d_{h}d_{j,h}}\right) \right). 
%\end{equation}
    \item A transition from $S_{1}$ to $S_{2}$ for an event $(i)$ occurs with probability {${P_{1,2}^{(j)} = \alpha p(d_{j}^{(i)})}$}. Otherwise, the IoTD stays in $S_1$ with probabilty ${P_{1,1}^{(j)} = \sum_{i=1}^{\text{ON time}} (1 - \alpha)^i \frac{\text{ON time}-i}{\text{DRX cycle}-i} }$, while transits from $S_{1}$ to $S_{4}$ with probabilty ${P_{1,4}^{(j)} = 1 - P_{1,1}^{(j)} - P_{1,2}^{(j)}}$. 
    \item A transition from $S_{2}$ to $S_{3}$ occurs when $B$ is enough to transmit, thus $P_{2,3}^{(j)} = \Pr(B\ge E_\text{Tx})$, while the device transits from $S_{2}$ to $S_{4}$, with probability $P_{2,4}^{(j)} = 1 - P_{2,3}^{(j)}$. 
    \item A transition from $S_{3}$ to $S_{1}$ occurs with probabilty {${P_{3,1}^{(j)} = \sum_{i=1}^{\text{On time}} \frac{\text{On time}-i}{\text{DRX cycle}-i}}$}, while the transition from $S_{3}$ to $S_{4}$ occurs with probabilty $P_{3,4}^{(j)} = 1 - P_{3,1}^{(j)}$. 
\end{enumerate}

\section{Optimization Framework}

An IoTD duty cycling and wake-up management policy is referred to as ${\boldsymbol{\delta}(k) = [\delta_1(k), \delta_2(k), \dots, \delta_N(k)]^\mathbf{T}}$ for each TTI $k \in [0,K]$. At each TTI $k$, %the BS assigns a value of 1 or 0 to each 
$\boldsymbol{\delta}(k)$ %, which 
designates which IoTD will be in $S_1$, waiting to be triggered or waken-up from $S_4$ to $S_2$ (designated by 1), and which device will remain in $S_4$ (designated by 0).   
Here, $\boldsymbol{\delta}(k)$ must take into account ${\Pr(B^{(j)}\ge E_\text{idle})}$ and ${\Pr(B^{(j)}\ge E_\text{Tx})}$ due to the energy arrivals at random times and the finite battery storage capacity. %Moreover, due to the finite battery storage capacity, the energy level in the battery never exceeds $E_{\text{max}}$. Since energy arrives at certain time points, it is sufficient to ensure that the energy level in the battery never exceeds $E_{\text{max}}$ at the times of energy arrivals. 

%When the online time ($K$) is sufficiently long, 
By considering a long time interval (\textit{i.e.,} large $K$), 
minimizing the probability of missing events %becomes a challenge of minimizing energy consumption. 
mainly depends on the availability of energy in the battery of the IoTDs. 
Reducing energy consumption means having a greater density of IoTDs with high energy availability for event sensing and reporting. This in turn leads to fewer blind spots where events may go unnoticed, consequently decreasing the misdetection probability.  
Then, we formulate the %problem to solve. Specifically, the 
optimization problem as 
%\begin{equation}
        %\text{P}1: \text{ } \max_{\boldsymbol{\delta}(k)} \text{ } \frac{1}{K}\sum_{\forall {(i)}} I^{{(i)}},   
        %\text{P}1: \text{ } \min_{\boldsymbol{\delta}(k)} \text{ } \frac{1}{K}\sum_{I^{{(i)}}} P_{\text{miss}}, 
%\label{P1}
%\end{equation}
\begin{subequations}
\begin{alignat}{4}
        \text{P}1: \text{ } &\min _{\boldsymbol{\delta}(k)} 
        \text{ } 
        &&\overline{\text{EC}}\\
        &\text{ } \text{s.t.} &&\frac{1}{\alpha K}\sum_{\forall {(i)}} I^{(i)} \geq  I_{\min}, 
        %& &&\mathbb{E}(P_{\text{miss}}) &&\leq \text{  } M,
\end{alignat}
\label{P1}
\end{subequations}
\hspace{-2mm}where %$(i)$ denotes event occurrence and 
$I_{\min}$ is the minimum expected information per event. Here, we use the expected number of events, $\alpha K$, as normalization factor instead of the exact number of events since the latter cannot be known.  Herein, %the misdetection error ($P_{\text{miss}}$) is calculated as the proportion of events that are not detected. 
$\overline{\text{EC}}$ can be reformulated from~\eqref{EC} as 
\begin{align}
    \overline{\text{EC}} = \frac{1}{NK}{\sum\limits_{k=1}^{{K}} 
    \sum\limits_{j=1}^{N} {\delta_j}(k) 
    (E_{\text{idle}} + \Pr(S_3^{(j)}) E_{\text{Tx}})}. 
    \label{EC2}
\end{align}
Note that at each TTI $k$, IoTDs with $\delta_j = 0$ consume no energy in sleep mode $S_4$ while IoTDs with $\delta_j = 1$ need high energy availability for sensing/transmit. Therefore, proper $\boldsymbol{\delta}(k)$ management is crucial at each TTI.    
%where $\boldsymbol{\delta}(k)$ represents the BS management policy for each IoTD, 
%where 
%\begin{equation}
    %\begin{array}{rl}
        %\delta_j &= \begin{cases}
        %1, &\text{for IoTDs in }S_1,
        %\\
        %\hfil 0,   &\text{for IoTDs in }S_4.\\
    %\end{cases}\\
    %\end{array}
%\end{equation}

%Inherently, solving problem P1 in~\eqref{P1} entails enhancing energy efficiency within the network. This is because minimizing the misdetection error necessitates maintaining the battery level of IoTDs above the threshold required for transmission. Only these devices have the capability to transmit event information. Thus, by minimizing the misdetected events, the energy efficiency of the network is also maximized. Herein, we also maintain the level of information that the BS requires from an event above a threshold ($I_{\min}$), having access to more information or a certain level of information from IoTDs enhances the BS's ability to effectively manage the network, respond to events, and ensure the overall reliability and performance of the IoT system.   

%Inherently, solving problem P1 in~\eqref{P1} entails enhancing minimizing the misdetection error. 
Satisfying a minimum expected information per event necessitates maintaining IoTDs' sensing/reporting availability. That is maintaining the battery level of IoTDs above the threshold required for transmission.  Only these devices have the capability to transmit event information. At the same time, proper sensing cycling and wake-up strategy helps to reduce energy consumption and avoid misdetections, which worsen the mean information per event. %Herein, we maintain the level of information that the BS requires from an event above a threshold ($I_{\min}$), having access to more information or a certain level of information from IoTDs enhances the BS's ability to effectively manage the network, respond to events, and ensure the overall reliability and performance of the IoT system.

%\subsection{Heuristic \& KNN-based Duty Cycling Optimization}\label{proposal}

Optimization methods to solve this problem can be both computationally intensive and time-consuming, particularly in the context of large-scale IoT systems where the complexity of performing interior point method can increase significantly, scaling with the number of IoTDs as $\mathcal{O}(N^{\rho})$, with $\rho \in[3,4]$~\cite{colombo2007advances}. 
%Given these challenges, optimizing decision-making algorithms to minimize computational complexity is crucial for enabling energy-efficient and self-sustainable IoT networks. 
In light of this, we aim to propose heuristic and data-driven approaches to address the issue in a more computationally efficient manner. Heuristic approaches, while potentially providing near-optimal solutions within a short time-frame, rely on intuitive rules or strategies. %On the other hand, data-driven solutions leverage available data to learn patterns and make informed predictions.

\subsection{Heuristic \& KNN-based Duty Cycling Optimization}\label{proposal}

Herein, we outline the potential approaches to tackle the problem at hand. The proposed approach is summarized
in Algorithm~\ref{alg1}.  
%Here we present Algorithm~\ref{alg1} outlining potential approaches to address the problem.
%It's worth noting that heuristic learning can be conducted online without the need for a comprehensive network overview for re-training. Specifically, sensors can autonomously implement and adapt strategies in response to re-training events, utilizing packet acknowledgments as feedback while being aware of their fixed positions within clusters and their respective thresholds.
Specifically, we propose two approaches. 
In the first one, we optimize the IoTD duty cycling by using an exhaustive search within the ON time ($S_1$) and DRX cycle\footnote{The duration of ON time ($S_1$) plus sleep time ($S_4$) within a cycle constitutes the DRX cycle.} values. 
In the second one, we form clusters in line 2 of Algorithm~\ref{alg1} based on each device's spatial placement. We  create a graph structure using KNN method as follows: {{(1) set cluster number randomly, (2) calculate the distance to the neighbors, (3) identify each IoTD KNN based on the distances, and (4) assign each IoTD to the cluster that the majority of its KNN belong to, and (5) update centroids and repeat (2-5) until convergence~\cite{liu2011output}.}}
The algorithm sets a maximum distance $d_{\text{max}}$ in line 1, which corresponds to the minimum sensing power $p(d_{\text{max}})$ set to form the clusters. The aim is to reduce the misdetection probability by limiting the sensing area that the clusters must control.
%Thus each data point is connected to its cluster's neighbors, creating a graph structure. 
Then, at line 3, we configure the duty cycling of the IoTDs in each cluster. 
%During each TTI within a cluster, the IoTDs rotate so that in each TTI, one IoTD is actively sensing in $S_1$ while the others are in $S_4$, ensuring that there is always at least one IoTD sensing in $S_1$ at each TTI. 
In each TTI, the IoTDs within a cluster iterate the  sensing in $S_1$. That is, during each TTI, only one IoTD is actively sensing in $S_1$ while the other IoTDs in the cluster are in $S_4$, ensuring there is always one IoTD in $S_1$ during each TTI. In each cluster, the DRX cycle duration for IoTDs within that cluster is equal to the number of IoTDs in that cluster.  
%Additionally, when an event occurs, 

In both approaches, the BS may request more information about an event from the other devices in $S_4$ that have strong spatial correlation with the IoTDs in $S_3$, as in line 7. In this way, the BS just requests the information that is needed using the WuR technology. 
Herein, we maintain the level of information that the BS requires from an event above a threshold ($I_{\min}$), having access to more information or a certain level of information from IoTDs enhances the BS's ability to effectively manage the network, respond to events, and ensure the overall reliability and performance of the IoT system. 
%As stated in Section~III, we propose equipping the IoTDs with WuR technology. This allows the BS to have access to all IoTDs in each TTI, even when the IoTDs are in a low-power consumption state ($S_4$). This ensures that the IoTDs remain available while consuming minimal power in $S_4$. 
%Algorithm~\ref{alg1} summarizes the proposal approach in this section where $d_{\text{max}}$ (line 1) is established to delimit a minimum sensing power $p(d_{\text{max}})$ to form the clusters. Herein, we pretend to decrease the misdetection probability by restricting the sensing area that the clusters have to control. 
Note that a shorter DRX cycle decreases the misdetection probability when there is high energy availability, however, it increases the energy consumption and the risk of low energy availability. Then, a trade-off between energy saving and sensing availability to lower the misdetection probability is paramount.

\begin{algorithm}[t]
\caption{KNN-based solution}\label{alg1}
\begin{algorithmic}[1]
%\vspace{-2mm}
\State \textbf{Set} number of clusters ($M$) as $M = \xi/(\pi d_{\text{max}}^2)$
\State \textbf{Form} clusters using KNN method 
\State \textbf{Configure} duty cycling of IoTDs according to its cluster association (ON time = 1, duty cycle = cluster size)
    \For{each $k$} 
        \If{BS receives information from active IoTDs}  
        \State $\hat{I}^{(i)} = \max(I_{h}^{(i)}), \forall h \text{ in } S_3$
            \If{$\hat{I}^{{(i)}} < I_{\min}$}
                \State {BS sequentially activates IoTDs ${\forall j \text{ in } S_4}$ with ${\Pr\left( S_3^{(j)}|S_3^{(h)} \right) \geq p(d_{\text{max}})}$, ${\forall h \text{ in } S_3}$, starting with the one with the highest value while updating ${\hat{I}^{{(i)}} = \max(\hat{I}^{{(i)}},I_{j}^{(i)}), \forall j \text{ in } S_3}$}
            \EndIf 
        \State \textbf{Update} $I^{{(i)}} \leftarrow \hat{I}^{{(i)}}$
        \EndIf
    \EndFor
\end{algorithmic}
\end{algorithm}
%\small{$^2${(1) Set cluster number randomly, (2) calculate the distance to the neighbors, (3) identify each IoTD KNN based on the distances, and (4) assign each IoTD to the cluster that the majority of its KNN belong to, (5) update centroids and repeat (2-5) until convergence~\cite{liu2011output}.}}

%In this paper, we propose equipping IoTDs with a wake-up radio (WuR) modules to detect requests from the BS to either enter the active state ($S_2$) or switch between activation states. The WuR module allows IoTDs with low-energy availability to remain in state $S_4$ while also ensuring their availability to provide additional information to the BS when required ($P_{4,2}$).
%The BS may try to wake up certain IoTDs in $S_4$ to sense potential events based on information received from other IoTDs and their spatial correlation. However, this might not always be possible because IoTDs with low-energy availability stay in $S_4$ (with probability $P_{4,4}$) and harvest energy for future events (if enough energy becomes available). Meanwhile, IoTDs in $S_4$ with high-energy availability may be authorized by the BS to move from $S_4$ to $S_1$ for the purpose of monitoring potential events.  

%In the context of IoT, the network remains stationary, ensuring that the devices' positions are fixed and can be determined offline. 
%In this paper, we use KNN to form the clusters. 

Notice that the first approach differentiates from the second one in the way the duty cycling is configured. Indeed, the BS wake-up request for the first approach is according to the device's correlation rather than cluster-based as in Algorithm~\ref{alg1}.

\subsection{Complexity Analysis}

The first approach of optimizing the duty cycling involves an exhaustive search within the ON time and DRX cycle. However, this method has a significant computational cost, with a complexity of $\mathcal{O}(g^N)$. Here, $g$ is equal to the number of possible ON time values multiplied by the number of possible DRX cycle values. On the other hand, the second approach implementing KNN is bounded to $\mathcal{O}({N}({N}-1)/2)$. Such a worst-case complexity arises due to the distance computation to the ${N}$ IoTD for each query point using a brute-force approach. The search involves scanning through the entire dataset to identify the KNN, resulting in a linear time complexity. However, as previously mentioned, the computation time can differ based on the utilized algorithm, occasionally reducing to $\mathcal{O}(\min \{M({N} - M), ({N} - M)^2\})$, where $M$ is the number of clusters~\cite{liu2011output}. %Meanwhile, an exhaustive search within the ON time and DRX cycle entails a complexity of $\mathcal{O}()$   

%\subsection{Performance Metrics}

%In this section, we discuss two performance metrics, named energy consumption and misdetection error, which are utilized to assess the performance of the proposed method and to compare it with the benchmark. %Additionally, ..

%\subsection{Energy Consumption}

%\subsection{Energy Consumption}
%Moreover, the misdetection error is calculated as the proportion of events that are not detected. 

\section{Simulations}\label{result}
  
%Then, let's deploy ${N}$ devices in 
We consider 
a 20$\times$20 m$^2$ area with ${N} \in %[10^{-2} 10^{-1}]$ devices/m$^2$ (between 25 and 250 devices) 
[10 \text{ } 250]$ devices
%and try to maximize the received information at the BS while minimizing energy consumption. 
since ${N} < 10$ is not enough to cover the area, provoking many misdetection events. 
%Additionally, $\Pr(A_j | A_h) = 0.56\times e^{-0.73d_{j,h}}$ from fitting the values in~\cite{ul2022learning}, and assuming $\eta = 1$~\cite{ruiz2022energy}. 
We perform $10^3$ runs of Monte Carlo simulations with different deployments in each run and $10^4$ TTIs, while assuming $\eta = 1$~\cite{ruiz2022energy} and $I_{{\min}} = \exp{(-2)}$. In addition, we assume $E_\text{Tx} = 10$ units while $E_\text{idle} = 1$ unit,  and $p(d_{\text{max}})=0.018$ corresponding to $d_{\text{max}} = 4$ meters~\cite{ul2022learning}. We assume IoTDs experiencing energy harvesting processes with the same statistics, \textit{i.e.}, $\lambda_j = \lambda, \forall j \in \mathcal{N}$. 
%We also assume that the BS is able to know or at least predict the battery level of the devices according to the information received from those IoTDs in $S_4$. By adopting this approach, a device with low-energy availability will not be prompted to wake up and transmit information. Such an action would drain the battery and, even if the device were due to transmit, the information would not be received due to the lack of energy required for transmission.   

%\clearpage
\subsection{Benchmarks}

We consider two benchmarks. In the first one, we let each device to adopt a random duty cycling with ON time $\in [1,2]$ and DRX cycle $\in [2, 4, 8]$. 
For the second one, we adopt a genie-aided approach wherein the closest IoTD to the event epicenter always detects the event. It is important to mention that this approach is not feasible in practice. This is because the BS would need to know in advance which IoTD is the closest to the event epicenter in order to activate it using WuR. However, this approach provides a reference lower bound.

\subsection{Performance Comparison}

\begin{figure}[t]
	\centering
	\includegraphics[width=0.9\columnwidth]{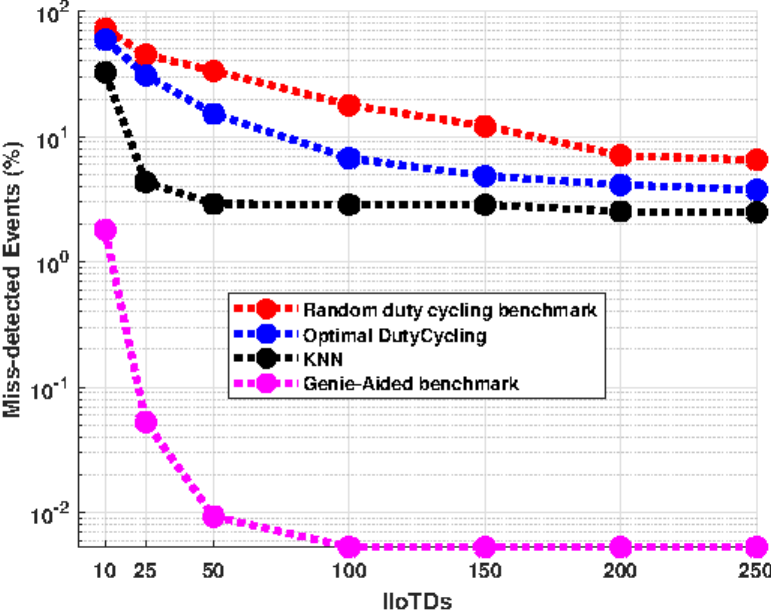}
 \vspace{-3mm}
    \caption{Percentage of misdetected events.}
	\label{miss_fig}
 \vspace{-3mm}
\end{figure}

%We are conducting an evaluation of our proposed methods for different deployment configurations. 
In Fig.~\ref{miss_fig}, we present the misdetection probability for each configuration as a function of the device density. %We have used the proposed cluster-based solution with spatial correlation, the duty cycling solution with spatial correlation, and a benchmark for comparison. 
It is worth noting that the misdetection probability decreases as the device density increases. For instance, when there are only 10 IoTDs, the misdetection probability is around 33\% for the cluster-based proposal, while for the duty cycling proposal is 58\%, and 77\% for the random duty cycling benchmark. The reason for this is that there might be blind spots in the deployment, that is no IoTDs near certain event occurrences. However, as the IoTD density increases, the probability drops below 3\% for the proposed KNN-based solution. For the random duty cycling benchmark, the misdetection probability drops to a minimum of 7\% for 250 IoTDs. %, but the performance is low due to low-battery availability in the IoTDs. 
%It is important to note that the intelligent management of the network and IoTD duty cycling is crucial for sustainable massive IoT deployment. 
Notice that the KNN-based solution outperforms both the optimal duty cycling proposal and the random duty cycling benchmark. Even though the optimal duty cycling can outperform the benchmar by up to a 200\%, just optimizing the duty cycling is not enough to lower the misdetection probability below 4\% in high density scenarios. Note that with the genie-aided approach, the misdetection probability drops from 2\% to near 0.005\% as we increase the IoTD density. This is because IoTDs have higher energy availability, as they do not need to waste energy on sensing while there is no event happening. However, there is still a small probability that the IoTD that detected the event does not have enough energy to transmit, as shown by this 0.005\%.    

\begin{figure}[t]
	\centering
	\includegraphics[width=0.9\columnwidth]{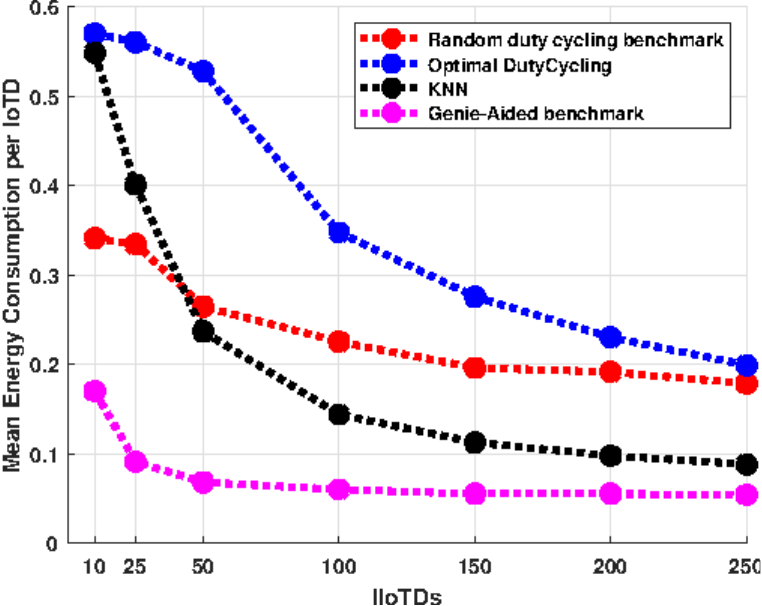}
  \vspace{-3mm}
    \caption{Mean Energy consumption per device per TTI.}
	\label{EC_fig}
 \vspace{-3mm}
\end{figure}

Hereinafter, we also investigate the mean energy consumption per device per TTI as the density of IoTDs %using the proposal method and how lowering the misdetection probability afects both proposals
increases. The results depicted in Fig.~\ref{EC_fig} show that the mean energy consumption for the proposal methods is around 0.55-0.57 units per IoTD per TTI, while for the random duty cycling benchmark, it is around 0.34 units for 10 IoTDs. 
However, this consumption is according to the random duty cycling benchmark poor performance in detecting the events.  
As the device density increases, the mean energy  consumption decreases up to 0.088 units for the proposed KNN-based method. In contrast, the mean energy consumption for the random duty cycling benchmark is above 0.18 units per TTI, which means that energy consumption increases by more than 100\%. Notably, for more than 40 IoTDs, the mean energy consumption of the proposed KNN-based method is lower than for the random duty cycling benchmark. Additionally, our proposed method shows better performance, as seen in Fig.~\ref{miss_fig}, where the misdetection probability is 11 times lower. It is worth-noting that the lower misdetection probability in the duty cycling proposal translates into a higher energy consumption for low-density scenarios. However, this gap disappears as the IoTD density increases. Meanwhile, compared to the genie-aided approach, the energy consumption to avoid misdetection of events is up to three times more for low density scenarios when using the KNN-based proposal. However, as the IoTD density increases, using the KNN approach can help reduce the gap in energy consumption. In high density scenarios, this approach can result in a reduction of up to 0.088 units per IoTD per TTI, compared to 0.055 units per IoTD with the genie-aided approach. This translates to just a 39\% increase in energy consumption compared with the genie-aided.   

\begin{figure}[t]
	\centering
	\includegraphics[width=0.9\columnwidth]{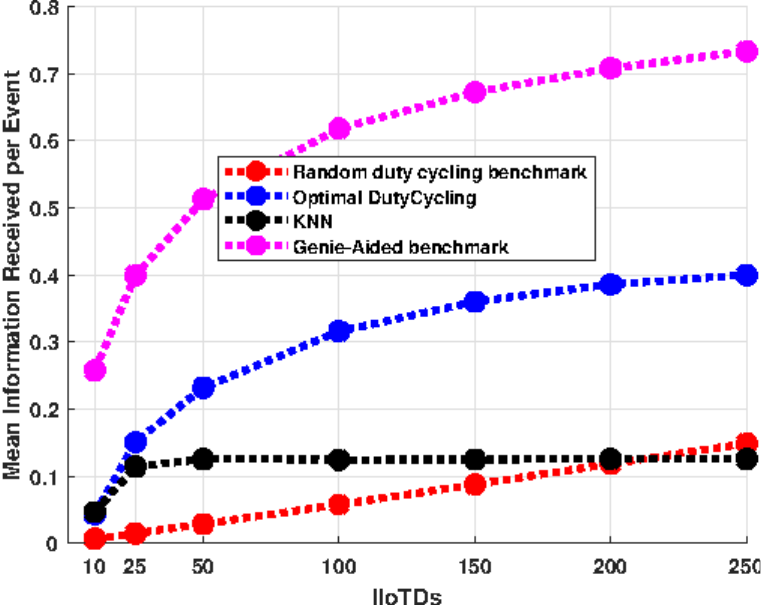}
  \vspace{-3mm}
    \caption{Mean information received by the BS per event.}
	\label{fig5}
  \vspace{-3mm}
\end{figure}

%Fig.~\ref{fig5} shows the average information received by the BS per event. Note that we established $I_{{\min}} = \exp{(-2)}$ and the proposed KNN-based method converges around this value for more than 25 IoTDs while the benchmark is not able to reach this $I_{{\min}}$ for values below 200 IoTDs. Meanwhile, the information provided by the duty cycling proposal increases with the IoTD density, providing more information since more IoTDs are active at the same time, giving more information than the BS requests and increasing the mean energy consumption. Note that with the genie-aided method, the BS always obtains more than the minimum requested information from the event, the information increases with the IoTD density. 
In Fig.~\ref{fig5}, we show the average information received by the BS per event. It is worth noting that we have established $I_{{\min}} = \exp{(-2)}$ and that the proposed KNN-based method converges around this value for more than 25 IoTDs, whereas the random duty cycling benchmark is not able to reach this $I_{{\min}}$ for values below 200 IoTDs. On the other hand, the information provided by the duty cycling proposal increases with the density of IoTDs. This is because more IoTDs are active at the same time, providing more information than the BS requests and increasing the mean energy consumption. Note that with the genie-aided method, the BS always obtains more than the minimum requested information from the event.

It is essential to note that using the proposed KNN-based method requires a higher computational cost. However, the proposed method runs on the BS, and the cost for the IoTDs is negligible. Moreover, the KNN-based proposal can be performed offline with the spatial information of the IoTDs. %Additionally, the approach described can further enhance its performance in scenarios exhibiting characteristics resembling clusters. In this particular case, the devices were randomly distributed.

\section{Conclusions}\label{conclusion}

IoT sustainable development requires comprehensive strategies integrating energy-efficient technologies and  optimized resource allocation methods. This paper focused on efficiently managing IoT resources by extending device battery life and addressing low-energy availability. We  modeled an IoT network with energy harvesting capabilities, employing a four-state Markov chain to depict IoTDs' states and modulated Poisson processes for energy harvesting availability. Additionally, the battery state was modeled as a discrete-time Markov chain. 
%We formulated the optimization problem to minimize the mean energy consumption at the IoTDs and proposed duty cycling configurations with a wake-up method using WuR at the IoTDs, considering energy efficiency trade-offs and spatial-temporal correlations between IoTDs. 
We proposed duty cycling configurations for IoTDs and using a wake-up method with WuR to minimize mean energy consumption, while considering spatial-temporal correlations and energy efficiency trade-offs.
The results demonstrated significant improvements in energy savings and performance compared with a random duty cycling benchmark, with up to 11 times less misdetection probability and a more than 50\% reduction in energy consumption observed in high-density scenarios.

%\bibliographystyle{IEEEtran}
%\bibliography{bib}

% Generated by IEEEtran.bst, version: 1.14 (2015/08/26)

\end{document}